\documentclass[pdflatex,sn-mathphys-num]{sn-jnl}

\usepackage{amsmath,amssymb,amsfonts}
\usepackage{mathtools}
\usepackage{bm}
\usepackage{graphicx}

\numberwithin{equation}{section}

\begin{document}

\title[Interpretation of Bell Violations]
{Contextuality, Locality, and the Interpretation of Bell Violations}

\author*[1]{\fnm{Partha} \sur{Ghose}}\email{partha.ghose@gmail.com}

\affil*[1]{%
\orgname{Tagore Centre for Natural Sciences and Philosophy},
\city{Kolkata},
\postcode{700156},
\country{India}}

\abstract{
The ontological significance of violations of Bell inequalities is
reexamined. Bell's theorem admits several mathematically equivalent
formulations that support different interpretive perspectives.
The present paper develops one such perspective using contextuality,
the sheaf-theoretic framework of Abramsky and Brandenburger,
and Nelson's stochastic mechanics.
}


\maketitle

\section{Introduction}

Since the publication of the seminal paper of Einstein, Podolsky, and Rosen (EPR) \cite{EPR1935}, the question of whether quantum mechanics implies nonlocality has remained central to foundational debates. Following Bell's theorem \cite{Bell1964} and the experimental violations of Bell inequalities in entangled quantum systems, it has become common to assert that nature itself is nonlocal. This conclusion, though widely accepted, is not logically unavoidable. It rests on one possible interpretation of Bell's result rather than being uniquely entailed by it. The purpose of the present paper is to develop an alternative interpretive framework based on Fine's theorem \cite{Fine1982}, the sheaf-theoretic approach to contextuality of Abramsky and Brandenburger \cite{Abramsky2011}, and Nelson's stochastic mechanics \cite{Nelson1966,Nelson1985,GuerraMorato1983}. Within this framework, Bell violations are understood as manifestations of contextuality rather than as compelling evidence for nonlocality.

\section{The EPR Argument Revisited}
The EPR argument begins with a locality or spatial separability assumption: the real state of a system should not depend on operations performed on a spatially separated system. On this basis, EPR introduce their criterion of reality:\begin{quote}If, without in any way disturbing a system, we can predict with certainty the value of a physical quantity, then there exists an element of physical reality corresponding to that quantity.\end{quote}Applying this to entangled states, EPR argue that one may choose to measure different observables on one system and thereby predict corresponding values on the distant  non-ineracting system. From this, they infer that multiple observables must simultaneously correspond to elements of reality. Since quantum mechanics does not assign simultaneous values to such observables, they conclude that it is incomplete.The critical step is \emph{the transition from alternative possible measurements to simultaneous reality}. This step involves an implicit counterfactual assumption.

\subsection{Einstein's Later Dissatisfaction with the EPR Paper}

It is important, historically and conceptually, not to identify the published
EPR paper too closely with Einstein's own mature formulation of the issue.
Although the argument arose from discussions among Einstein, Podolsky, and
Rosen, Einstein later made clear that the published form did not express the
point as simply as he had wished. In a letter to Schr\"odinger dated 19 June
1935, Einstein wrote that the paper had been written by Podolsky, for reasons
of language, after extensive discussions, but that the result had not come out
as he had wanted: the essential point, he said, had been ``buried by erudition''
\cite{EinsteinSchrodinger1935,Fine1996,Sauer2013}.

This remark is significant. It shows that Einstein did not regard the elaborate
EPR criterion of reality, nor the detailed discussion of simultaneous position
and momentum, as the deepest point of the argument. What mattered to him was a
simpler structural fact: after two systems have interacted and then separated,
the quantum state assigned to the second system can depend on which measurement
is chosen on the first. If the wave function is taken to be a complete
description of the real physical state of the second system, then the choice of
measurement on the first system appears to determine the real state of the
second, although the two systems are spatially separated and non - interacting.

Einstein stated this point with particular clarity in his \emph{Autobiographical
Notes} \cite{Einstein1949}. Let $S_1$ and $S_2$ be two systems that interacted and then became
spatially separated. Quantum mechanics allows one, by choosing two alternative
measurements on $S_1$, to assign two different wave functions to $S_2$. But Einstein
insisted on the following locality--separability requirement:
\begin{quote}
But on one supposition we should, in my opinion, absolutely hold fast: the real
factual situation of the system $S_2$ is independent of what is done with the
system $S_1$, which is spatially separated from the former.
\end{quote}
On this basis he concluded that the wave function cannot be in one-to-one
correspondence with the real physical state of an individual system.

Einstein's mature argument may hence be stated as follows. First, quantum
mechanics assigns different state functions to $S_2$ depending on the
measurement performed on the distant system $S_1$. Secondly, locality requires
that the real factual situation of $S_2$ cannot depend on this distant
experimental choice. Therefore, the quantum state assigned to $S_2$ cannot be a
complete description of its real physical state. The conclusion is not, in the
first instance, that nature is nonlocal, but rather that the quantum-mechanical
state description is incomplete if locality holds.

This formulation is sharper than the published EPR argument. The latter is
often read as resting on the simultaneous reality of non-commuting observables.
Einstein's later formulation shifts the emphasis from the possession of
simultaneous values to the relation between the wave function and the real `factual' state
of an individual system. The central issue becomes whether one and the same real
state of $S_2$ can be represented by different quantum states, depending on a
freely chosen measurement performed on the distant system $S_1$.

Thus Einstein's final position presents a clear dilemma:
\[
\text{either the wave function is complete and locality fails,}
\]
or
\[
\text{locality is retained and the wave function is incomplete.}
\]
As we will see, Bell's theorem \cite{Bell1964} sharpened this dilemma further by showing that locality,
measurement independence, and a global assignment of values to all possible
measurement outcomes cannot jointly reproduce the quantum correlations. The
important point for the present discussion is that Einstein's mature argument
was not primarily an argument for nonlocality. It was an argument that the
completeness of the quantum state is incompatible with the locality--separability
principle.

\section{Bohr's Response to the EPR Paper and the\\ Role of Context}

Bohr rejected the EPR criterion of reality by emphasizing the non-separability of physical quantities from the experimental arrangements used to measure them \cite{Bohr1935}. In modern terms, his position can be expressed as follows:\begin{quote}Physical quantities are defined only within specific measurement contexts, and statements drawn from \emph{mutually exclusive} contexts cannot be combined into a single description of reality.\end{quote} Thus, Bohr denies the legitimacy of the counterfactual step in the original EPR argument. Importantly, he does \emph{not} invoke nonlocality; rather, he challenges the assumption of \emph{context-independent reality}.

\section{Bell's Theorem and Its Assumptions}

There is substantial disagreement in the literature concerning the logical status of determinism or global value assignment in Bell's theorem. Bell himself regarded deterministic response functions as a consequence of locality together with the EPR correlations and maintained that determinism is inferred rather than assumed, a position developed further by Norsen and defended in different ways by Maudlin and by Wiseman \cite{Bell2004,Norsen2017,Norsen2006,Maudlin2011,Wiseman2014}. 

The present paper does not dispute Bell's historical reconstruction. Rather, it adopts the mathematically equivalent characterization provided by Fine's theorem \cite{Fine1996} and the sheaf-theoretic framework of Abramsky and Brandenburger \cite{Abramsky2011}, and explores the ontological interpretation suggested by that formulation. 
 
\section{Commuting Observables and Experimental\\ Contexts}

In a typical Clauser–Horne–Shimony–Holt (CHSH) \cite{CHSH1969} experiment, which extends Bell's theorem to imperfect correlations, only one pair of observables is measured per run--
one setting on Alice's side (say $A$ or $A^\prime$) and one setting on Bob's side (say $B$ or $B^\prime$)-- and one measures a pair like ($A,B$), ($A,B^\prime$), ($A^\prime, B$) or ($A^\prime, B^\prime$).
Each such pair commutes:
\[
[A,B]=[A,B^\prime] = [A^\prime,B] = [A^\prime,B^\prime] = 0.
\] 
because these observables are associated with spatially separated systems. Thus, there is no contradiction at the level of any single measurement. The difficulty arises when one attempts to combine results from different experimental contexts. 

From the perspective adopted here, following Fine's theorem \cite{Fine1996} and the sheaf-theoretic framework of Abramsky and Brandenburger \cite{Abramsky2011}, CHSH inequalities may be interpreted as expressing the existence of a single joint assignment of values (or equivalently a global probability distribution) for all observables appearing in different measurement contexts.
This is not operationally justified when the observables are locally incompatible, like different spin/polarization components on Alice's or Bob's side:  
\[
[A, A^\prime] \neq 0, [B, B^\prime] \neq 0.
\]
 
\subsection{Contextuality and the Failure of Global Assignment}

Modern formulations of contextuality, particularly the sheaf-theoretic approach of Abramsky and Brandenburger \cite{Abramsky2011}, provide a natural framework for analyzing the issues raised above. The essential idea can be understood without invoking the full mathematical machinery.

In any experiment of the Bell–CHSH type, one considers several distinct experimental arrangements or \emph{contexts}, corresponding to different choices of jointly measurable observables. For each such context, the experiment yields a well-defined probability distribution of outcomes. For example, in a CHSH experiment, one measures correlations for the four settings $(A,B)$, $(A,B')$, $(A',B)$, and $(A',B')$, each defining a separate experimental context.

The crucial question is whether these context-dependent probability distributions can be regarded as arising from a single underlying description. More precisely, can one construct a single joint probability distribution for all observables $A, A', B, B'$ whose marginals reproduce the experimentally observed distributions in each context? Such a joint distribution would correspond to a global assignment of values to all observables, independent of which measurements are actually performed.

Bell–CHSH inequalities express precisely the constraints that any such global assignment must satisfy. Their experimental violation therefore shows that no such joint distribution exists for entangled quantum states. In other words, the observed statistics cannot be embedded into a single, context-independent probabilistic model.

This is what is meant by \emph{contextuality}: the outcome statistics depend irreducibly on the measurement context, and cannot be explained by pre-existing values assigned independently of that context. The failure of a global assignment thus reflects a structural limitation on how measurement outcomes can be consistently represented, rather than indicating any dynamical influence propagating between spatially separated systems.

\section{A Stochastic-Mechanical Perspective on Entanglement and Measurement}
Before concluding, it would be useful to briefly examine how these issues appear in an alternative formulation of quantum mechanics due to Nelson \cite{Nelson1966,Nelson1985,GuerraMorato1983}. In this approach, quantum mechanics is reformulated in terms of stochastic processes in configuration space, with the wave function written in polar form
\begin{equation}
\psi(q,t)=R(q,t)\exp\!\left(\frac{i}{\hbar}S(q,t)\right),
\end{equation}
where $q=(x_1,x_2)$ denotes the joint configuration of a two-particle system. The probability density is $\rho=R^2$, and the `current velocity' $v(q,t)$ and `osmotic velocity' $u(q,t)$ are related to $S(q,t)$ and $\rho$  by
\begin{equation}
v(q,t)=\frac{1}{m}\nabla S(q,t), \qquad u(q,t)=\nu\,\nabla \ln \rho(q,t),
\end{equation}
where $\nu=\hbar/2m$ and $\nabla=(\nabla_{x_1},\nabla_{x_2})$.  The stochastic dynamics is then described by forward and backward diffusion
processes. In Nelson's notation these may be written as
\begin{align}
dX_t &= b(X_t,t)\,dt + \sqrt{2\nu}\,dW_t, \qquad dt>0,\\
dX_t &= b_*(X_t,t)\,dt + \sqrt{2\nu}\,dW^*_t, \qquad dt<0,
\end{align}
where \(W_t\) and \(W_t^*\) are Wiener processes. The forward and backward
drifts are related to the current and osmotic velocities by
\begin{equation}
v=\frac{1}{2}(b+b_*), \qquad u=\frac{1}{2}(b-b_*),
\end{equation}
or equivalently
\begin{equation}
b=v+u, \qquad b_*=v-u.
\end{equation}
Thus the drift fields of the stochastic process are determined by the phase
\(S\) and the density \(\rho=R^2\) of the wave function.

For the two-particle case, one writes $X(t) = (X_1(t), X_2(t))$ so that the SDE is on the full configuration space.

For entangled states, the functions $S(q,t)$ and $\rho(q,t)$ are non-separable functions of the joint configuration $(x_1,x_2)$. As a result, the velocities associated with one particle depend on the position of the other:
\begin{equation}
v_1=v_1(x_1,x_2,t), \qquad u_1=u_1(x_1,x_2,t),
\end{equation}
and similarly for particle 2. This reflects the fact that the stochastic process is defined on the full configuration space, rather than on independent single-particle spaces. In this sense, the correlations characteristic of entangled states are already built into the joint probability structure.

Consider now a measurement performed on particle 1. In the stochastic-mechanical framework, such a measurement can be naturally described as a conditioning of the joint probability distribution. If the outcome of a position measurement on particle 1 is $x_1=y$, the joint distribution $\rho(x_1,x_2)$ is updated to the conditional distribution
\begin{equation}
\rho^{(y)}(x_2)=\rho(x_2\mid x_1=y)=\frac{\rho(y,x_2)}{\int \rho(y,x_2)\,dx_2}.
\end{equation}
Correspondingly, one may define an updated wave function for particle 2 by
\begin{equation}
\psi^{(y)}(x_2)\propto \psi(y,x_2),
\end{equation}
from which new current and osmotic velocities for particle 2 can be computed.

Two distinct descriptions must be clearly distinguished. If one does not condition on a particular outcome of the measurement on particle 1, the marginal distribution for particle 2,
\begin{equation}
\rho_2(x_2)=\int \rho(x_1,x_2)\,dx_1,
\end{equation}
remains unchanged. In this unconditioned description, the stochastic dynamics of particle 2 is unaffected by the distant measurement, and no signalling is possible. On the other hand, if one conditions on a specific outcome $y$, the distribution $\rho^{(y)}(x_2)$ changes, and so do the associated velocities. This change is instantaneous in the sense that it reflects an update of the probabilistic description, but it is purely inferential in character, arising from conditioning rather than from a dynamical influence propagating between the two systems.

In the stochastic-mechanical framework, such a measurement can be naturally described as a Bayesian conditioning of the joint probability distribution, as discussed by Pavon \cite{Pavon1989}. The correlations between the two subsystems are already encoded in the joint distribution, and measurement serves to select a sub-ensemble corresponding to the observed outcome.

This analysis reinforces the viewpoint developed in the preceding sections. The nontrivial feature of entangled systems is not the presence of a dynamical influence acting instantaneously at a distance, but the impossibility of representing the observed correlations by a single, context-independent assignment of values to all observables. The stochastic-mechanical framework makes explicit that the apparent nonlocal features arise from the structure of the joint probability distribution and its conditioning, rather than from any signal propagating between spatially separated systems.

\subsection{A simple Gaussian example}

To illustrate the above discussion, consider a two-particle Gaussian state of the form
\begin{equation}
\psi(x_1,x_2)=N \exp\!\left[-\frac{(x_1-x_2)^2}{4\sigma^2}
-\frac{(x_1+x_2)^2}{4\Sigma^2}\right],
\end{equation}
where $N$ is a normalization constant. This state is entangled whenever $\sigma \neq \Sigma$. For simplicity, we take the phase to vanish so that the current velocity is zero, $v=0$, and all nontrivial structure is contained in the osmotic velocity.

The joint probability density is
\begin{equation}
\rho(x_1,x_2)=|\psi(x_1,x_2)|^2
= N^2 \exp\!\left[-\frac{(x_1-x_2)^2}{2\sigma^2}
-\frac{(x_1+x_2)^2}{2\Sigma^2}\right].
\end{equation}
The osmotic velocity of particle 2 is
\begin{equation}
u_2(x_1,x_2)=\nu\,\partial_{x_2}\ln \rho(x_1,x_2),
\end{equation}
which gives explicitly
\begin{equation}
u_2(x_1,x_2)=\nu\left[\frac{x_1-x_2}{\sigma^2}
-\frac{x_1+x_2}{\Sigma^2}\right].
\end{equation}
Thus the velocity of particle 2 depends explicitly on the position of particle 1, reflecting the non-separable character of the state.

Now consider a position measurement on particle 1 yielding the outcome $x_1=y$. The conditional probability density for particle 2 becomes
\begin{equation}
\rho^{(y)}(x_2)\propto \exp\!\left[-\frac{(y-x_2)^2}{2\sigma^2}
-\frac{(y+x_2)^2}{2\Sigma^2}\right].
\end{equation}
From this we obtain the updated osmotic velocity
\begin{equation}
u^{(y)}_2(x_2)=\nu\,\partial_{x_2}\ln \rho^{(y)}(x_2)
=\nu\left[\frac{y-x_2}{\sigma^2}
-\frac{y+x_2}{\Sigma^2}\right].
\end{equation}
Two features are noteworthy. First, the unconditional marginal distribution
\begin{equation}
\rho_2(x_2)=\int \rho(x_1,x_2)\,dx_1
\end{equation}
is independent of whether a measurement is performed on particle 1, so that no signalling occurs. Secondly, conditioning on a specific outcome $y$ leads to an immediate change in the effective dynamics of particle 2 through the updated velocity field. This change, however, arises from the conditioning of a joint probability distribution rather than from a dynamical influence propagating between the two particles.

This simple example makes explicit that the correlations of an entangled state are already encoded in the joint probability structure, and that the apparent ``collapse'' associated with measurement corresponds to a Bayesian update of that structure.

\section{Conclusion}
The analysis presented above has emphasized that the violation of Bell--CHSH
inequalities does not, by itself, compel the conclusion that nature is nonlocal.
Rather, such violations demonstrate the impossibility of embedding the observed
context-dependent statistics into a single, context-independent global
description. 

In much of the literature, the assumption that is abandoned is locality, leading
to the widespread conclusion that quantum mechanics entails nonlocality.
However, another mathematically equivalent and interpretive possibility is to retain locality while rejecting
the assumption of a global assignment of values, i.e.\ counterfactual
definiteness. From this perspective, Bell violations are naturally understood as
manifestations of contextuality.

This viewpoint is increasingly reflected in contemporary discussions. Kupczy\'nski has,
for example, argued that the interpretation of Bell violations in
terms of nonlocality is not obligatory, and that a contextual description
provides a viable and conceptually coherent alternative \cite{Kupczynski2023}.
Such analyses reinforce the idea that the experimental results constrain the
structure of admissible models, but do not uniquely determine their ontological
interpretation.

At the same time, it is important to recognize that contextuality and nonlocality
are closely related notions. In particular, it has been shown that, under suitable
conditions, contextual correlations can be transformed into nonlocal ones
\cite{Cabello2021}. This highlights a deep structural connection between the two
concepts. However, this connection should not obscure the logical distinction
emphasized in the present work. The existence of such mappings does not imply
that Bell--CHSH violations, taken in their original operational setting,
necessarily require a nonlocal causal interpretation.

The central lesson is therefore one of logical clarity. Bell's theorem does not
establish nonlocality in isolation; it establishes the incompatibility of a
specific set of assumptions. The widespread identification of Bell violations
with nonlocality arises from privileging one particular resolution of this
incompatibility. An alternative resolution, which retains locality and abandons
global value assignment, leads naturally to a contextual interpretation.

In this sense, the present analysis aligns with Bohr's insistence on the
contextual character of physical quantities, while incorporating the precision
afforded by modern formulations of contextuality. The challenge is not to decide
between locality and nonlocality in isolation, but to recognize that the classical
notion of a single, context-independent description of reality is no longer
tenable. 

The issue, therefore, is not whether nature is nonlocal, but whether our classical expectations about the existence of a global, context-independent reality can be sustained in the light of quantum phenomena.

\section{Acknowledgements}
The author acknowledges the use of ChatGPT for language polishing and assistance with manuscript preparation. The scientific content, interpretations, and conclusions are entirely those of the author.

\end{document}